# Investigating exchange, structural disorder and restriction in Gray Matter via water and metabolites diffusivity and kurtosis time-dependence

Eloïse MOUGEL[1], Julien VALETTE[1], Marco PALOMBO[2,3]

[1] *Université Paris-Saclay, Commissariat à l'Energie Atomique et aux Energies Alternatives (CEA), Centre National de la Recherche Scientifique (CNRS), Molecular Imaging Research Center (MIRCen), Laboratoire des Maladies Neurodégénératives, 92260 Fontenay-aux-Roses, France*

[2] *Cardiff University Brain Research Imaging Centre (CUBRIC), School of Psychology, Cardiff University, Cardiff, UK*

[3] *School of Computer Science and Informatics, Cardiff University, Cardiff, UK*

Corresponding author: Dr. Marco Palombo, Cardiff University Brain Research Imaging Centre, School of Psychology and School of Computer Science and Informatics, Cardiff University, United Kingdom. E-mail: palombom@cardiff.ac.uk.

Present/permanent address: Cardiff University Brain Research Imaging Centre, School of Psychology and School of Computer Science and Informatics, Cardiff University, United Kingdom.

**Abstract:**

Water diffusion-weighted MRI is a very powerful tool for probing tissue microstructure, but disentangling the contribution of compartment-specific structural disorder from cellular restriction and inter-compartment exchange remains an open challenge.

In this work we use diffusion-weighted MR spectroscopy (dMRS) of water and metabolite as a function of diffusion time *in vivo* in mouse gray matter to shed light on: i) which of these concomitant mechanisms (structural disorder, restriction and exchange) dominates the MR measurements and ii) with which specific signature.

We report the diffusion time-dependence of water with excellent SNR conditions as provided by dMRS, up to a very long diffusion time (500 ms). Water kurtosis decreases with increasing diffusion time, showing the concomitant influence of both structural disorder and exchange. However, despite the excellent experimental conditions, we were not able to clearly identify the nature of the structural disorder (i.e. 1D *versus* 2D/3D short-range disorder). Measurements of purely intracellular metabolites diffusion time-dependence (up to 500 ms) show opposite behavior to water, with metabolites kurtosis increasing as a function of diffusion time. We show that this is a signature of diffusion restricted in the intracellular space, from which cellular microstructural features such as soma's and cell projections' size can be estimated. Finally, by comparing water and metabolite diffusion time dependencies, we attempt to disentangle the effect of intra/extracellular exchange and structural disorder of the extracellular space (both impacting water diffusion only). Our results suggest a relatively short intra/extracellular exchange time (~1-50 ms) and short-range disorder (still unclear if 1D or 2D/3D) most likely coming from the extracellular compartment.

This work provides novel insights to help interpret water diffusion-time dependent measurements in terms of the underlying microstructure of gray matter and suggests that diffusion-time dependent measurements of intracellular metabolites may offer a new way to quantify microstructural restrictions in gray matter.

**Abbreviations:**

CSF: cerebrospinal fluid

**Graphical abstract:**

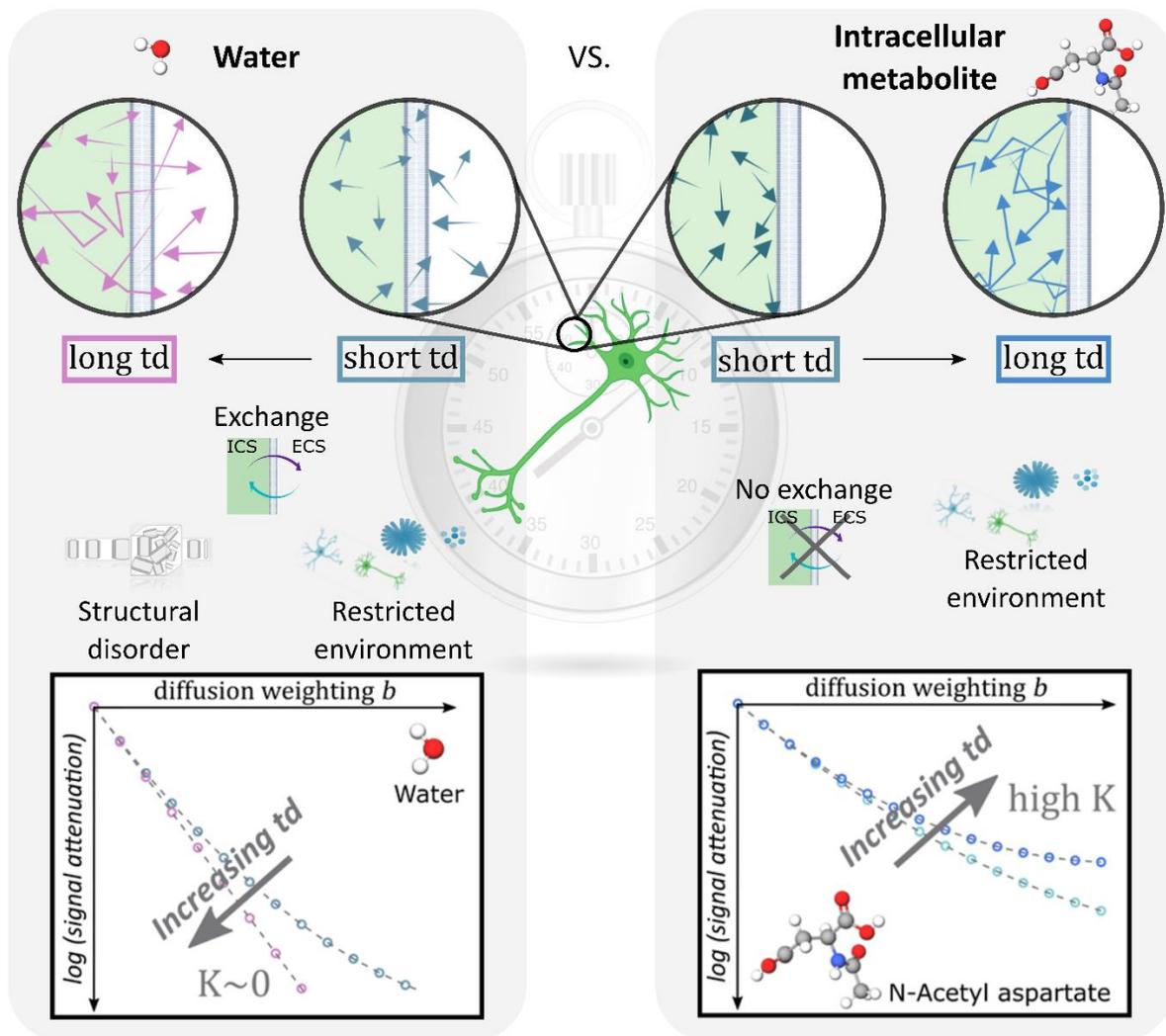

**Highlights:**

- Revisit water diffusion time-dependence with excellent SNR conditions as provided by dMRS and up to very long $t_d$ (500 ms)

- Report for the first time intracellular metabolite diffusion time-dependence up to very long $t_d$ (500 ms) to determine the signature of intracellular diffusion

- Compare water and metabolite diffusion time dependence to try to disentangle the effect of exchange and the signature of the extracellular diffusion.

# 1. Introduction

Diffusion-weighted MRI (dMRI) is a valuable radiological tool to non-invasively quantify the brain structure at the cellular scale, the so-called *microstructure* (Alexander et al., 2019; Jelescu et al., 2020; Novikov et al., 2019). Among the different dMRI approaches investigated in the past decades, measurements of the diffusion-time ($t_d$) dependent dMRI signal have been shown to provide information on the brain tissue inter-compartmental exchange (Hill et al., 2021; Jelescu et al., 2022; Nedjati-Gilani et al., 2017; Nilsson et al., 2009; Olesen et al., 2022; Stanisz et al., 1997; Zhang et al., 2021), structural disorder (Novikov et al., 2014; Palombo et al., 2018a, 2013) and the restricted environment where the molecules diffuse over ~1-100 µm (Assaf and Cohen, 2000; Does et al., 2003; Drobnjak et al., 2010; Jespersen et al., 2018; Palombo et al., 2016; Panagiotaki et al., 2012).

In particular, it has been shown that measurements of the $t_d$-dependence of water diffusivity (Aggarwal et al., 2012; Arbabi et al., 2020; Baron and Beaulieu, 2014; Does et al., 2003; Wu et al., 2014), $D_W(t_d)$, and kurtosis (Aggarwal et al., 2020; Lee et al., 2020; Pyatigorskaya et al., 2014; Wu et al., 2018), $K_W(t_d)$, provide unique quantitative insight into these tissue properties. Varying the $t_d$ allows probing different length scales and thus assessing the spatial heterogeneity of the cellular microenvironment. For instance, at short $t_d$, obtained with the oscillating gradient spin echo (OGSE) sequence, a large $D_W$ variation as a function of the oscillating-gradient frequency (inversely proportional to $t_d$) was measured in cell-dense regions compared to other regions of the mouse brain, which is thought to reflect, among other things, restricted diffusion on a length scale of a few micrometers, thus in the order of the cell nucleus size (Aggarwal et al., 2012). At longer $t_d$, which can be probed using, for example, pulsed gradient spin echo (PGSE), the length scales probed are more on the order of tens of micrometers. For such $t_d$, $D_W$ and $K_W$ are sensitive to demyelination in white matter (Aggarwal et al., 2020) or to 1D structural disorders associated with neurites in gray matter (GM) (Lee et al., 2020; Novikov et al., 2014), but it has also been shown to be influenced by exchange (Jelescu et al., 2022; Li et al., 2017; Olesen et al., 2022; Yang et al., 2018), and disentangling the different contributions remains complex. Most importantly, some exchange regimes can also influence $K_W(t_d)$, as shown by numerical simulations (Aggarwal et al., 2020): increasing the diffusion time will result in increased $K_W$ until $t_d$ approaches a characteristic time corresponding to a diffusion length comparable to the typical (permeable) restriction size; for longer $t_d$, $K_W$ will then decrease due to exchange, with a slope which is steeper as exchange gets faster (i.e. higher permeability). However, as valuable as these simulations are, they do not fully mirror the true structural complexity of the GM tissue, in particular the structural disorder of the intra- and extracellular space. Disentangling compartment-specific structural disorder from inter-compartment exchange, either experimentally or in simulations, remains very challenging.

In contrast and complementary to water, diffusion measurements of *purely intracellular* metabolites by diffusion-weighted MRS (dMRS) offer information about the brain microstructure that is specific to the intracellular space and is not affected by inter-compartmental exchange (Assaf and Cohen, 1999; Palombo et al., 2018b; Ronen and Valette, 2015). Some brain metabolites are also preferentially compartmentalized within neurons, such as N-acetyl aspartate (NAA), and glial cells, such as myo-inositol (Ins), providing a probe of neuronal or glial structure, and can, for example, be used to estimate neurite radii and intracellular diffusivity (Ligneul et al., 2017; Palombo et al., 2018b). Interestingly, $t_d$-dependent dMRS is an efficient tool for probing complex GM cell microstructure, as it was shown that dMRS is sensitive for example to spines and leaflets density (Palombo et al., 2018a) and cellular processes branching (Palombo et al., 2016). Specifically, $t_d$-dependence of metabolite diffusivity $D_M(t_d)$ and kurtosis $K_M(t_d)$ is expected to be influenced by features of the cellular structure at different length scales, but without influence of extracellular space or exchange with extracellular space (Ianus et al., 2021). Micrometric restriction (soma and projection size) or exchanges between soma and projection (Ianus et al., 2021) could be all specific information of the intracellular environment. $D_M(t_d)$ and $K_M(t_d)$ would thus provide valuable and complementary information to water to investigate GM. By comparing the $t_d$-dependent diffusivity, $D_M(t_d)$, and kurtosis, $K_M(t_d)$, of intracellular metabolites with $D_W(t_d)$-$K_W(t_d)$ one can potentially separate and quantify the relevant mechanism(s) driving each $t_d$-dependency in GM.

The aim of the present work is to shed some light on the role of exchange, structural disorder and restriction in GM, by exploiting the complementary information of $D_W(t_d)$-$K_W(t_d)$ and $D_M(t_d)$-$K_M(t_d)$ measured by dMRS. More specifically, it aims at:

(I) revisit water diffusion time-dependence $D_W(t_d)$-$K_W(t_d)$ with excellent SNR conditions as provided by dMRS (i.e. measured in a large volume) and up to very long $t_d$ (500 ms)

(II) report for the first time, to our knowledge, intracellular metabolite diffusion time-dependence $D_M(t_d)$-$K_M(t_d)$ up to very long $t_d$ (500 ms) to determine the signature of intracellular diffusion

(III) combining results from (I) and (II) to try to disentangle the effect of exchange and the signature of the extracellular diffusion.

## 2. Material and methods

### 2.1. dMRS acquisition

#### 2.1.1. Ethics approval, scanner and sequence

All experimental protocols were reviewed and approved by the local ethics committee (CETEA N°44), and authorized by the French Ministry of Education and Research. They were performed in an approved facility (authorization #B92-032-02), in strict accordance with recommendations of the European Union (2010-63/EEC). All efforts were made to minimize animal suffering, and animal care was supervised by veterinarians and animal technicians. Mice were housed under standard environmental conditions (12-hour light-dark cycle, temperature: 22 ± 1 °C and humidity: 50%) with *ad libitum* access to food and water.

Experiments were performed on an 11.7 T BioSpec Bruker scanner interfaced to PV6.0.1 (Bruker, Ettlingen, Germany). A quadrature surface cryoprobe (Bruker, Ettlingen, Germany) was used for transmission and reception. Wild-type C57BL/6 mice were anesthetized with ~1.6% isoflurane and maintained on a stereotaxic bed with one bite and two ear bars. Throughout the experiment, body temperature was monitored and maintained at ~36 °C by warm water circulation. Breathing frequency was monitored using PC – SAM software (Small Animal Instruments, Inc., Stony Brook, NY). The sequence used for all acquisition was a diffusion-weighted stimulated-echo sequence (Ligneul et al., 2017) followed by a LASER localization (STE-LASER). Echo time was set to TE = 33.4 ms (including the 25-ms LASER echo time) and was held constant for each acquisition. Repetition time was set to TR = 2500 ms. The duration of the pulsed diffusion gradient was set to δ = 3 ms. Diffusion gradients were separated by the delay Δ, which was varied according to the protocol described in the next two sections. In addition, a VAPOR water-suppression module supplemented by an additional 21-

ms hermite inversion pulse inserted during the mixing time was used for metabolite acquisition, respectively.

*2.1.2. Water acquisition*

Water diffusion was measured in four mice. A spectroscopic volume of interest (1.5 × 0.8 × 2 mm³ = 2.4 µL) was placed in the hippocampus, in such a way that it consisted almost exclusively of gray matter (no visible white matter, and CSF contamination <<1%). For each delay = 21.8, 31, 43.5, 101, 251, 501 ms i.e. resulting respectively in diffusion time $t_d$ = 20.8, 30, 42.5, 100, 250, 500 ms, attenuation was measured for $b$ = 0.2, 0.7, 1.2, 2.0, 2.5 ms/µm² along one fixed direction, with 16 repetitions.

*2.1.3. Intracellular metabolites acquisition*

Metabolite diffusion was measured in seven mice. A spectroscopic voxel (7 × 1.5 × 3 mm³ = 31.5 µL) covering the two hemispheres was centered in the hippocampus so that CSF contamination was minimal. Voxel composition (gray matter~94%, white matter~5% and CSF~1%) was estimated a posteriori on anatomical images acquired with a RARE sequence with TE/TR=30/2500 ms, 78.1-µm isotropic resolution and 0.5-mm slice thickness, using manual segmentation with the Fiji distribution of the ImageJ software, in the whole volume. For each delay = 43.5, 101, 251, 501 ms i.e. resulting respectively in diffusion time $t_d$ = 42.5, 100, 250, 500 ms, attenuations of N-acetylaspartate (NAA), choline compounds (tCho), creatine + phosphocreatine (tCr), inositol (Ins), taurine (Tau) were measured for $b$ = 0.2, 1.0, 2.0, 3.2, 4.5, 6.0, 8.0 ms/µm² along one direction. Two blocks of 32 repetitions interleaved between each $t_d$ and $b$ value were acquired for $t_d$ = 42.5, 100, 250 ms and three blocks of 32 repetitions were acquired to improve the SNR for $t_d$ = 500 ms.

*2.2 Data processing and analysis*

*2.2.1. Data processing*

Individual scans were frequency- and phased-corrected before averaging. For water acquisitions, the integral of water peak was computed on a 0.5-ppm window centered on the peak maximum at each *b* and $t_d$. For metabolites, each spectrum was analyzed with LCModel (Provencher, 2001). Experimental macromolecule (MM) spectra, acquired for each $t_d$ using a double inversion recovery module (TI$_1$ = 2200 ms and TI$_2$ = 770 ms), at *b* = 10 ms/µm², were included in LCModel basis-sets containing simulated metabolite spectra.

*2.2.2. Data Analysis*

Remaining within these predefined *b*-value range (according with the convergence radius of the cumulant expansion (Frøhlich et al., 2006) for one hand water and other hand metabolites, which exhibit about four times slower diffusivity than water), for each $t_d$, the diffusion weighted signal S as a function of *b* was fitted with a non linear least square method implemented in Matlab's 'lsqcurvefit' function to estimate the $t_d$-dependent apparent diffusivity of water or metabolite (D$_{W/M}$($t_d$)) and kurtosis (K$_{W/M}$($t_d$)), using equation 1 (Jensen and Helpern, 2003), for each $t_d$.

$$S(b) = S(0) \times exp(-bD + 1/6\ Kb^2D^2) \quad [1]$$

## 3. Results & discussion

### 3.1. Revisiting water diffusion time-dependence: which power law describes water $t_d$ dependence?

*Methodological considerations*

The STE-LASER sequence provides a good water signal even at a long diffusion time and a high *b* value, allowing D$_W$ and K$_W$ to be evaluated over a wide range of diffusion times [20.8; 500] ms and reaching a fairly long diffusion time ($t_d$~500 ms), longer than in previous studies using PGSE sequence (Jelescu et al., 2022; Lee et al., 2020). With respect to attenuation (Fig.1B), the curvature of signal attenuation clearly attenuates with increasing diffusion time, approaching but not reaching near mono-exponential diffusion at long diffusion time (K~0.3). The model given by equation [1] fits the attenuation of water well (as shown for one mouse on Fig.1B), and this protocol provides a small inter-animal distribution of diffusion parameters (D$_W$ and K$_W$), with a standard error of the mean between mice of 2% at shortest $t_d$ and ~15% at longest $t_d$ for D$_W$ and ~2% for K$_W$.

Regardless of the precise time-dependence, average $D_W$ estimated in this study are in the same order of magnitude as ADC measured in rat (Jelescu et al., 2022) and mouse GM (Aggarwal et al., 2020; Zhang et al., 2021), and slightly lower than the ADC in human brain (Lee et al., 2020). $K_W$ is also in the same order of magnitude as in previous studies (Jelescu et al., 2022; Lee et al., 2020). This spectroscopic acquisition using the STE-LASER sequence is therefore in good agreement with the literature.

### $D_W$ and $K_W$ decrease at increasing $t_d$

Both $D_W$ and $K_W$ are time dependent (Fig1.C-D), over the whole $t_d$ range. $D_W$ decreases significantly (~12% difference with an F-test ($F(5, 15) = 31.102$, $p < 0.001$) for a Repeated Measure ANOVA), reaches a plateau ~0.5 µm²/ms, when $D_W$ tends to the tortuosity asymptote. (Fig.1C). $K_W$ also decreases (~50% with an F-test ($F(5, 15) = 13.321$, $p < 0.001$) for a Repeated Measure ANOVA) and approaches ~0.3 (Fig.1D), as the diffusion time increases, consistent with the diffusion not being Gaussian even at the longest diffusion time probed in this experiment. Given the composition of the water spectroscopic voxel used (nearly totally GM), we are also confident that the presence of multiple $T_1$ pools, or exchange between multiple $T_1$ pools during TM, are very unlikely to bias our diffusion measurements. Furthermore, by plotting $\log S(b = 0) = f(TM)$ (not reported), TM being the mixing time, we did not observe multiple $T_1$ pools that could bias the signal, or diffusivity estimates as a function of $t_d$.

The main difference from previous results (Jelescu et al., 2022; Lee et al., 2020) is in the $D_W$ time dependence. On the one hand we measured a significant decrease in $D_W$ (from 0.62 to 0.57 µm²/ms) over the $t_d$ range 20-42.5 ms, while on the other hand (Jelescu et al., 2022) and (Lee et al., 2020) report an almost constant $D_W$ over this same $t_d$ range in rat cortex and hippocampus and over a $t_d$ range 20-100 ms in human cortical GM, respectively. On closer inspection, the ~0.05-µm²/ms difference measured in our study is within the range of inter-animal variability reported in these other studies, which may explain the lack of significant difference in these other studies.

A decrease in $D_W$ and $K_W$ over the entire $t_d$ range suggests a nontrivial influence of microstructure, such as structural irregularities, and therefore the diffusion cannot be considered Gaussian, even at the longest diffusion time measured in this experiment. Moreover, these observations may also indicate an influence of the exchange between the intracellular and extracellular spaces, or a contribution of the two phenomena, structural disorder and exchange.

*Structural disorder*

According to the literature (Lee et al., 2020; Novikov et al., 2014) intra-compartmental diffusion can be influenced by structural disorder, and thus the assumption of Gaussian diffusion is broken. In particular, works based on coarse-graining approach have shown that $D_W(t_d)$ and $K_W(t_d)$ can reflect different classes of structural disorder in 1D, 2D or 3D (Lee et al., 2020; Novikov et al., 2014). Therefore, we compared different functional forms to fit the data and identify which structural disorder might dominate the water behavior.

First, three models were applied to determine whether 1D or 2D/3D structural disorder could dominate. Based on (Lee et al., 2020; Novikov et al., 2014), we consider models given by

$$D(t_d) = D_\infty + C_D \, t_d^{-\theta} \quad [2]$$

$$K(t_d) = K_\infty + C_K \, t_d^{-\theta} \quad [3],$$

with the constants $C_D$, $C_K$, $D_\infty = \lim_{t_d \to \infty} D$ and $K_\infty = \lim_{t_d \to \infty} K$ left as free parameters to be determined by fitting the experimental data and setting the "universal" exponent $\theta = (p+d)/2$ to 0.5 (representing 1D structural disorder with a dimensionality d = 1 and an universality of disorder p = 0) and 1 (representing either 2D/3D structural disorder, with either p = 0 and d = 2 corresponding for example to random discs or p = -1 or d = 3 corresponding for example to random rods), respectively (Fig.2A-B). In gray matter, neuronal and glial cell soma, represented as spherical object with a radius >> sqrt(D0 $t_d$) ~ 5-8 µm, can also induce a 1/t power law decay of $D_W(t_d)/K_W(t_d)$. If water probes an extracellular space mostly comprised of randomly oriented fibers, $D_W(t_d)$ and $K_W(t_d)$ may also exhibit a logarithmic singularity of the form $\ln(t_d/t_c)/t_d$ at long $t_d$ (Burcaw et al., 2015). Hence, we also consider the alternative model

$$A \, (\ln(t_d/t_c))/t_d \; + B, \quad [4]$$

with A a coefficient proportional to the universal dimensionless tail ratio, and B a coefficient corresponding to $K_W$ in the long time limit (Fig.2C).

Visually, it is not clear whether $D_W$ and $K_W$ follow a power law with exponent $\theta = 0.5$, $\theta = 1$ or equation [4]. The results of the fit with the specific 1D structural disorder form are not too different from the results of the fit with $\ln(t_d/tc)/t_d$, according to cAIC that is low for both models. In addition, it is not clear that $D_W$ and $K_W$ follow exactly the same law. At this stage, it is therefore difficult to determine whether one or the other form of structural disorder dominates.

To go further, a model with exponent $\theta$ in equation [2] and equation [3] left as an additional free parameter to be determined from the data was used to fit $D_W$ and $K_W$ (Fig.3). The fit gives $\theta = 0.2 \pm 0.2$, (which is closer to 0.5 than 1), and the universal dimensionless tail ratio $\xi = C_K/(C_D/D_\infty) = 1.3 \pm 0.2$ (which is close to 2). Values of $\theta \sim 0.5$ and $\sim 2$ are expected for 1D short-range structural disorder (Lee et al., 2020; Novikov et al., 2014). We also found a tail ratio of ~2 regardless of the model used for fitting, with a best precision. Fit with free exponent is less accurate because of the small number of $t_d$.

### 3.2. Unique $t_d$ dependence of metabolite diffusion in the intracellular space

*Methodological considerations*

Spectra acquired in the hippocampal voxel have a good signal-to-noise ratio (>20 for NAA for the longest $t_d$ and highest $b$ value) even at the longest diffusion times and highest $b$ values (Fig.4) and the LCModel analysis (not shown here) is robust (CRLB ~2% for all metabolites even at the longest diffusion time and highest $b$ value) and we find a correct estimate of S/S0 with low inter-animal variability. The model given by equation 1 also fits the data well for each metabolite and animal (Fig.4).

*$D_M$ decreases and $K_M$ increases*

For each metabolite, with the exception of Ins, the inter-animal estimate of $D_M$ and $K_M$ is relatively consistent between animals with a standard error of the mean of less than 9% for $D_M$ and 30% for $K_M$ (Fig.5). $D_M$ decreases to a plateau of ~0.08 µm²/ms for NAA, Glu, tCr, and Tau, and ~0.06 µm²/ms for tCho and Ins with increasing $t_d$ for all metabolites, and the $D_M$ estimate is clearly in agreement with previous results (Palombo et al., 2016). Despite the low number of $t_d$, imposed by an in vivo experimental setup, statistical analysis with a paired sample t-test shows a significant increase in $K_M$ and decrease in $D_M$ for each metabolite, with at least a power of 0.7 for our sample size N=7. The diffusivity of predominantly intra-glial metabolites (Fisher et al., 2002) is lower than that of other metabolites. These differences are also noticeable in the estimate of $K_M$, which increases with $t_d$ for NAA, Glu, tCr, and Tau and tends almost to a plateau ~2.3-2.5 while Ins and tCho remain almost constant ~1.5.

Such an increase in K is expected for diffusion in a restricted environment and thus strongly affected by the microstructure, in agreement with the numerical simulations of (Ianus et al., 2021), which account for restriction in soma, exchange between soma and cellular processes and cellular processes branching. The differences on the diffusion parameters observed between glial and neuronal cells are therefore not surprising because of the rather different morphology of these two cell types (Palombo et al., 2018b). Besides the high sensitivity of these dMRS methods for probing microstructure, already reported in previous studies (Ligneul et al., 2019), this work shows that even at lower $b$ values ($\lesssim$10 ms/µm²), suitable for the validity range of the kurtosis representation, different cell microstructures can be distinguished. dMRS thus appears a powerful tool to probe the cell morphology with potential feasible translation to *in vivo* human measurements, using also clinical scanners with relatively low maximum gradient strength (~60-80 mT/m).

*Intracellular metabolites as endogenous probe of the intracellular microarchitecture*

To go further, we therefore proposed to evaluate the influence of the restricted environment on diffusion by fitting to $D_M(t_d)$ and $K_M(t_d)$ a simple model representing the intracellular compartment through the soma and projections formed by 20% spheres and 80% randomly oriented cylinders. Due to the low number of measurements, this ratio is imposed to avoid overfitting the model. Using an in-house iterative algorithm, the model was fitted to the $D_M$ and $K_M$ of mostly intra-neuronal (NAA and Glu) and predominantly intra-glial (tCho and Ins) metabolites (Fisher et al., 2002; Gill et al., 1989; Griffin et al., 2002; Urenjak et al., 1993), to extract radii from the soma and projections of (hypothetical) neuronal and glial cells, respectively (Fig.6). First, this restriction model appears to be suitable for describing the behavior of $D_M(t_d)$ and $K_M(t_d)$. For neuronal cells, this model gives estimates with low standard error of the radii of spheres and $D_M(t_d = 0)$, and the estimates are consistent with previous results, whereas the radii of cylinders are rather poorly estimated. For intra-glial metabolites, the uncertainty in the sphere and cylinder radii is high and the estimate is overestimated, but the $D_M(t_d = 0)$ is consistent with the literature in normal mouse GM (Ligneul et al., 2019; Palombo et al., 2020). Of course, the aim of this work is not to propose a new model for estimating biophysical parameters, but these results suggest that even such a simple model, like SANDI (Palombo et al., 2020), describes metabolites data well.

In conclusion, the $t_d$-dependent diffusion of intracellular metabolites is dominated by restriction in the intracellular space, in line with previous study with various dMRS method (Assaf and Cohen, 1999; Kroenke et al., 2004; Ligneul et al., 2017; Shemesh et al., 2017, 2014; Valette et al., 2018; Vincent et al., 2020) and it is not surprising that it is clearly influenced by cell morphology (cell projection size and soma size at least). Importantly, as $K_M$ does not decrease at increasing $t_d$, no signature of 1D structural disorder along the neurites is observed for the metabolites, in contrast with what has been speculated for intracellular water. The $t_d$ dependence of metabolite diffusion seems thus clearly more specific to cell morphology than water.

### 3.3. Intracellular metabolite diffusion to highlight the role of exchange in water diffusion time-dependence

It is insightful to examine water and metabolites results in the light of the work of Aggarwal *et al.* (Aggarwal et al., 2020). Our experimental results about intracellular metabolite kurtosis time-dependence, i.e. kurtosis increasing with $t_d$, are in line with the behavior simulated for long exchange times in (Aggarwal et al., 2020). In contrast, water kurtosis time-dependence exhibits the opposite behavior, which is in line with the behavior expected for a system with significant exchange in (Aggarwal et al., 2020). This strongly suggests fast water exchange relative to the diffusion times used in the present study, which is in agreement with previous works suggesting short exchange time (Jelescu et al., 2022; Lee et al., 2020). The influence of such fast exchanges is also suggested in our estimate of the free exponent, $\theta = 0.2$, found in Fig.3, which is close to that found by Jelescu *et al.* when they fitted their simulated exchange-driven K for a short exchange time ~25 ms, with the same power-law decay with a free theta exponent.

To provide a quantitative estimate of exchange time for our experimental conditions, we fit an increasingly used (Fieremans et al., 2010; Jelescu et al., 2022; Jensen, 2023; Li et al., 2023; Moutal et al., 2018; Olesen et al., 2022; Zhang et al., 2021) model of exchange: the Kärger Model

$$K(t_d) = K_0 \frac{2\,t_{ex}}{t_d} [1 - \frac{t_{ex}}{t_d}(1 - e^{-\frac{t_d}{t_{ex}}})] \quad [5]$$

$$K(t_d) = K_0 \frac{2\,t_{ex}}{t_d} [1 - \frac{t_{ex}}{t_d}(1 - e^{-\frac{t_d}{t_{ex}}})] + K_\infty \quad [6]$$

to the data for the longest diffusion time [42.5; 500] ms, to satisfy one of the assumptions of the Kärger model, i.e. constant $D_W(t_d)$ (Fig.7). The estimated exchange time is very imprecise and spans hundreds of ms: from 46 (±412) ms (using equation [6]) to 196 (±238) ms (using equation [5]). Regardless of the poor precision despite high data quality, the contradiction with the previous estimate of exchange time shorter than ~25 ms (as discussed in the previous paragraph) might be due to the fact that the assumptions behind the Kärger Model are not satisfied. According to Fieremans *et al.* (Fieremans et al., 2010), the non-applicability of the Kärger Model suggests that the membrane between the intracellular and extracellular spaces is too permeable to allow for an accurate description using the Kärger Model. Other recently proposed models of exchange are NEXI (Jelescu et al., 2022) and SMEX (Olesen et al., 2022); both based on the same modified Kärger Model to account for potential exchange between neurites and extracellular space. Fitting NEXI/SMEX to the water data as a function of *b* value (for extra *b* values up to 6 ms/µm²) and diffusion time (Fig.8) yields once more very imprecise estimates of exchange time: 1.7 (±4.9) ms (other model parameters estimates are: intraneurite diffusivity 1.56±1.28 µm²/ms; extraneurite diffusivity 0.75±1.10 µm²/ms and neurite signal fraction 0.64±0.52). Although imprecise, the NEXI/SMEX estimate of exchange time is more in agreement with the evidence of exchange time shorter than 10 ms discussed in the previous paragraph and supported by recent works (Cai et al., 2022; Olesen et al., 2022; Williamson et al., 2023, 2020, 2019).

In conclusion, this spectroscopic measurements with lower variability and over a wider range of $t_d$ than the previous studies (Jelescu et al., 2022; Olesen et al., 2022) corroborate the influence of exchange on water diffusion as a function of time (in vivo in the mouse GM). The comparison of the very different functional forms of the measured $D_W(t_d)$-$K_W(t_d)$ and purely intracellular $D_M(t_d)$-$K_M(t_d)$ supports the hypothesis that the signature of 1D or 2D/3D short range structural disorder observed for $D_W(t_d)$-$K_W(t_d)$ might mostly come from the water diffusion in the extracellular space, which "contaminates" the whole water pool signature due to fast exchange and hence hides intracellular signature. Our measurements reinforce the need for a model describing the $t_d$-dependence of diffusion in GM, with specific signatures in intra- and extracellular compartments and inter-compartmental exchange, to precisely evaluate how exchange can mask the signatures of intracellular space.

## 4. Limitations

The range of $t_d$ values over which we were able to carry out our measurements remains rather restricted, even if it is the widest ever probed in this type of *in vivo* measurement. On the one hand, a constraint linked to the duration of the spoiling gradient applied during the mixing time in this sequence (Ligneul et al., 2017), prevents us from having access to smaller $t_d$ values. With a shorter $t_d$ we could observe the inflection point of the time-dependent kurtosis of water, which could help determine the average permeability. Future work could try to combine OGSE and PGSE measurements to try to characterize the time dependence at shorter $t_d$. On the other hand, the increase in $t_d$ is limited by the SNR, which becomes too low beyond 500 ms. Among other things, this increase could allow us to better discriminate which class of structural disorder dominates, which would be a first step towards disentangling the different contributions of exchange and structural disorder.

Our aim here is not to propose a new modeling method, but to present experimental data that could support theoretical models to better describe gray matter. We have chosen to place ourselves in a diffusion regime suited to a representation with the cumulant exponent, which seems quite sensitive to exchange. Nevertheless, with our experiments we were not able to precisely determining the exchange time in gray matter (using currently available biophysical models, i.e. Kärger model and NEXI/SMEX), and to determine the influence of structural disorder on the water diffusion properties. The complexity of both exchange mechanisms (e.g. multiple exchange times) (Nilsson et al., 2013) and microstructure (e.g. glial and neuronal cells and complex extracellular space) (Palombo et al., 2020) will need to be taken into account in future work to disentangle the contributions of exchanges and structural disorder. Towards this goal, future work could harness advanced computational modeling tools for gray matter microstructure simulations (Callaghan et al., 2020; Palombo et al., 2020, 2016).

We also chose a smaller voxel size for water, in order to focus only on gray matter without multiple compartment effects that can complicate interpretation of the $t_d$ dependence of the diffusion and to be in conditions similar to those encountered in the literature for segmented images of rodent brain (Jelescu et al., 2022). In particular, in this voxel, we minimize the contribution of CSF and white matter. For metabolites, we have no choice to take this large voxel to have sufficient signal at long diffusion time, however the contribution of CSF and WM remains small (respectively ~5% and 1%), which can certainly have a negligible impact on our results. Indeed, we did not notice a strong signal attenuation at low $b$ which would have suggested a high CSF contribution whatever the metabolite. This is in line with measurements carried out by Lundell et al. (Lundell et al., 2021) in the human brain, where they noted that for a CSF contribution of ~2% there was no visible effect on diffusion, whereas for a contribution of ~10% there was. Finally, by fitting a bi-exponential model to log(S0) as a function of mixing time, only a small fraction of the short $T_1$ component was estimated, suggesting that signal is not affected by multi-$T_1$ pool, and certainly not influenced by short $T_1$ of WM.

In this work we focused on healthy mice; future work will explore the ability of combined $D_{W/M}(t_d)$-$K_{W/M}(t_d)$ to characterize pathological microstructural changes. Moreover, it could be informative to extend in future work the investigation of $D_M(t_d)$-$K_M(t_d)$ to also metabolites that are in both intra- and extra-cellular space and potentially in exchange. Further extension of this work will include measurements in humans to assess the feasibility and translatability of this approach.

## 5. Conclusion

This experimental study proposes for the first time a comparison of the $t_d$ dependence of the diffusion of two different endogenous probes, water and intracellular metabolites. The information provided by these two sets of $t_d$-dependent acquisitions is complementary, especially since they probe different compartments of the GM. In particular, the $t_d$ dependence of water diffusion is mainly dominated by short-range structural disorder in the extracellular space and rapid exchanges between intra- and extracellular spaces, whereas the $t_d$ dependence of intracellular metabolite diffusion is more specific to restriction within the intracellular space. Other underlying features could be further investigated to fully decipher the sources contributing to total kurtosis (beading, undulation, …), as already proposed by Shemesh Lab (Alves et al., 2022; Henriques et al., 2020). Nevertheless, these new results help interpret $D(t_d)$-$K(t_d)$ measurements in terms of the underlying microstructure of GM and suggest that $D(t_d)$-$K(t_d)$ measurements of metabolites may offer a new way to quantify microstructural restrictions in GM.

**Funding:** This project has received funding from the European Research Council (ERC) under the European Union's Horizon 2020 research and innovation programmes [grant agreement No 818266] and the UKRI Future Leaders Fellowship MR/T020296/2. The 11.7 T MRI scanner was funded by a grant from "Investissements d'Avenir - ANR-11-INBS-0011 - NeurATRIS: A Translational Research Infrastructure for Biotherapies in Neurosciences". M.P. is supported by the UKRI Future Leaders Fellowship MR/T020296/2.

# Figures

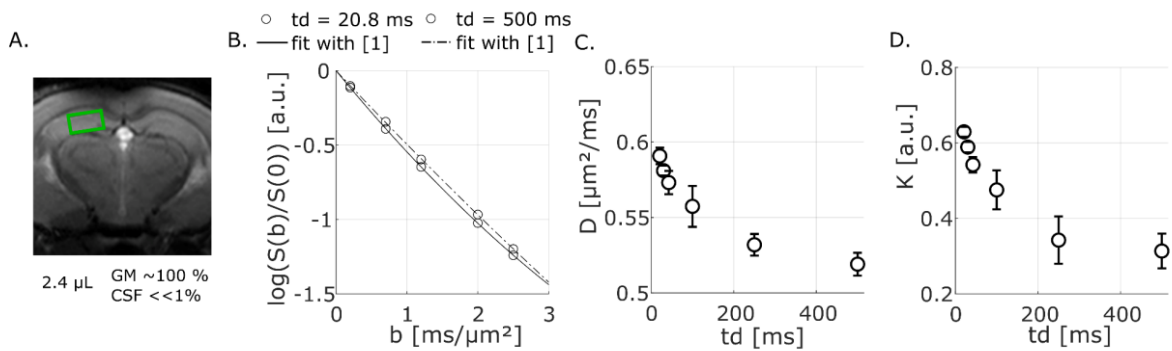

Fig.1: Water signal data. A. Voxel of interest, for water acquisition. B. Example of water signal attenuation acquired at two different $t_d$ on one mouse and fits with the kurtosis representation [1] (lines). C. Mean diffusivity and standard error of the mean over four mice. D. Mean kurtosis and standard error of the mean over four mice.

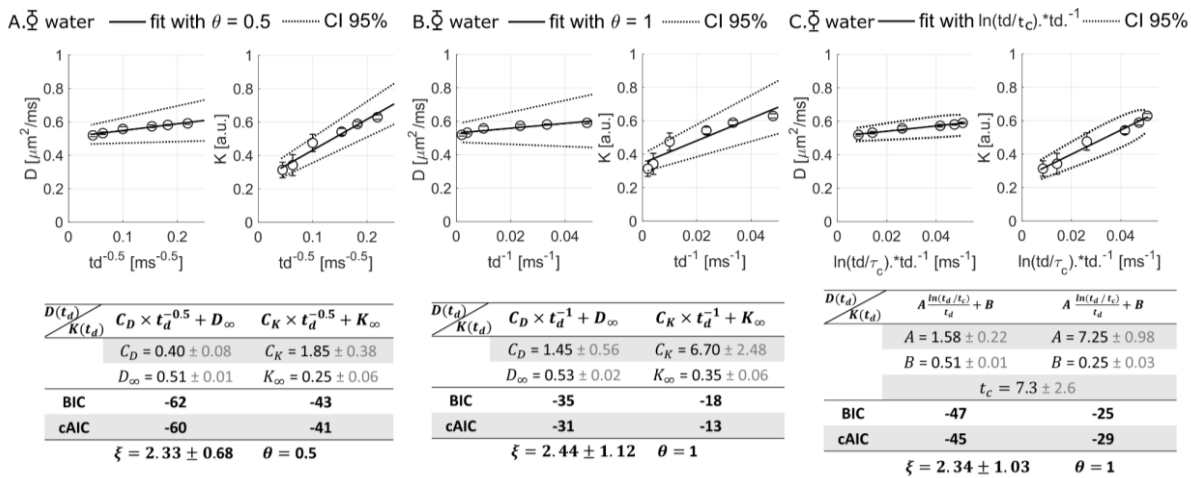

Fig.2: $D_W(t_d)$ and $K_W(t_d)$ averaged over four mice and different functional forms fitted to these data. A. The 1D structural disorder form is correctly fitted to D and K. B. The 2D structural disorder form does not fit well to the data, with the highest AIC for D and K. C. The 2D functional form in the limit of θ = (p+d) / 2 = 1 fits the data with a cAIC close to the cAIC estimate for the 1D structural disorder form. The dimensionless ratio $\xi = C_K/(C_D/D_\infty) \sim 2$ whatever the functional forms fitted to these data, which is compatible with pure 1D structural disorder ($\xi = 2$).

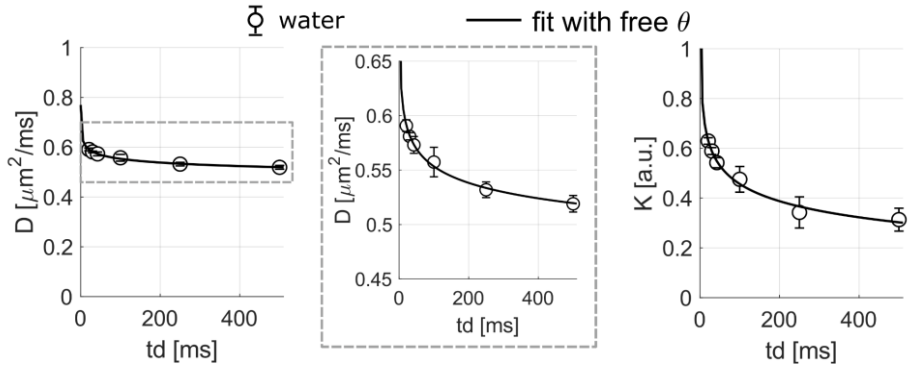

Fig.3: $D_W(t_d)$ and $K_W(t_d)$ averaged over four mice and fit with structural disorder model. Joint fit of a power-law decay with a free exponent theta to $D_W$ and $K_W$, with the same free exponent theta.

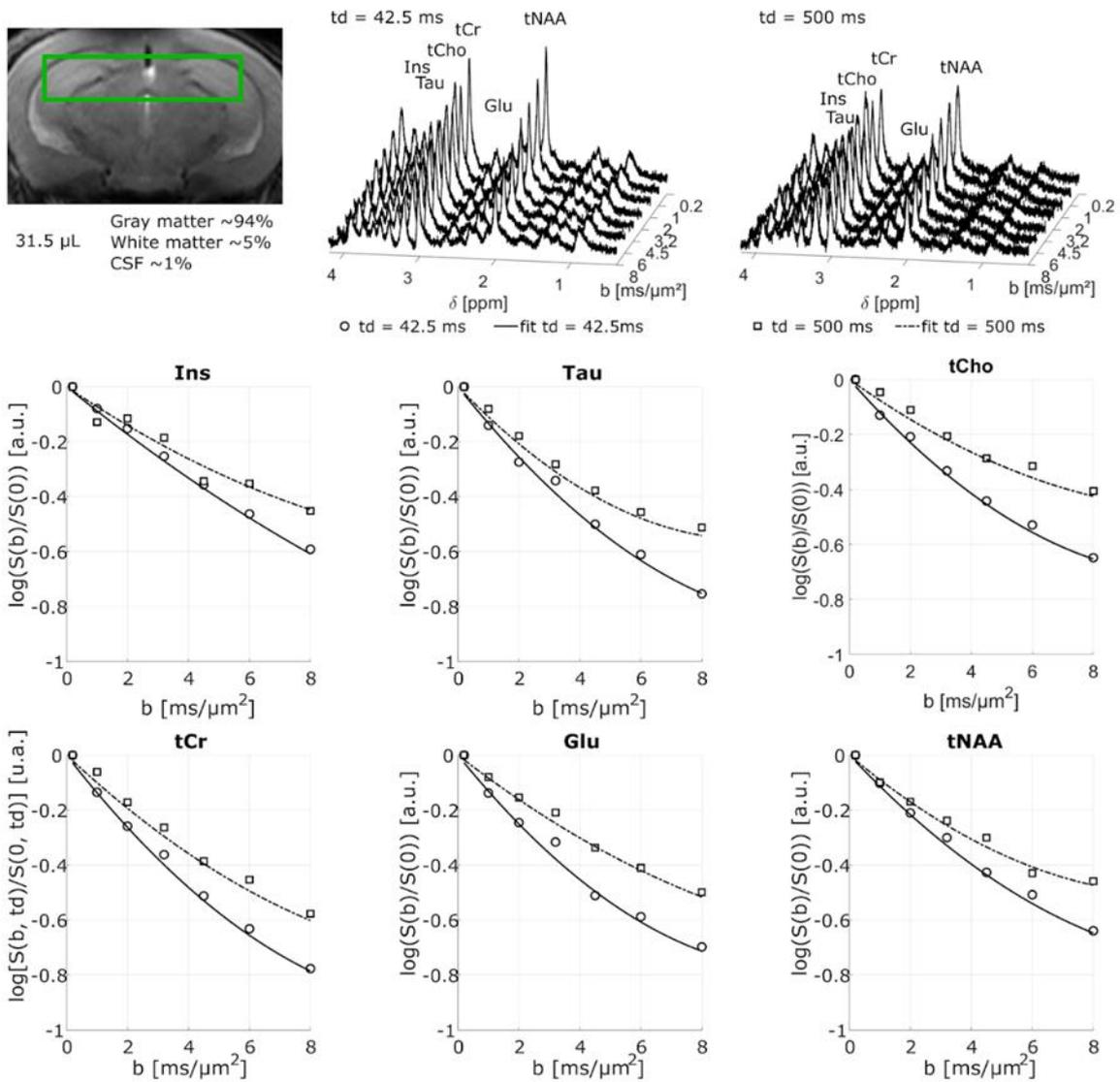

Fig.4: Example of spectra and attenuation curves for each metabolite as a function of diffusion-weighting $b$ acquired on one mouse for two different diffusion times in a voxel around hippocampus (image).

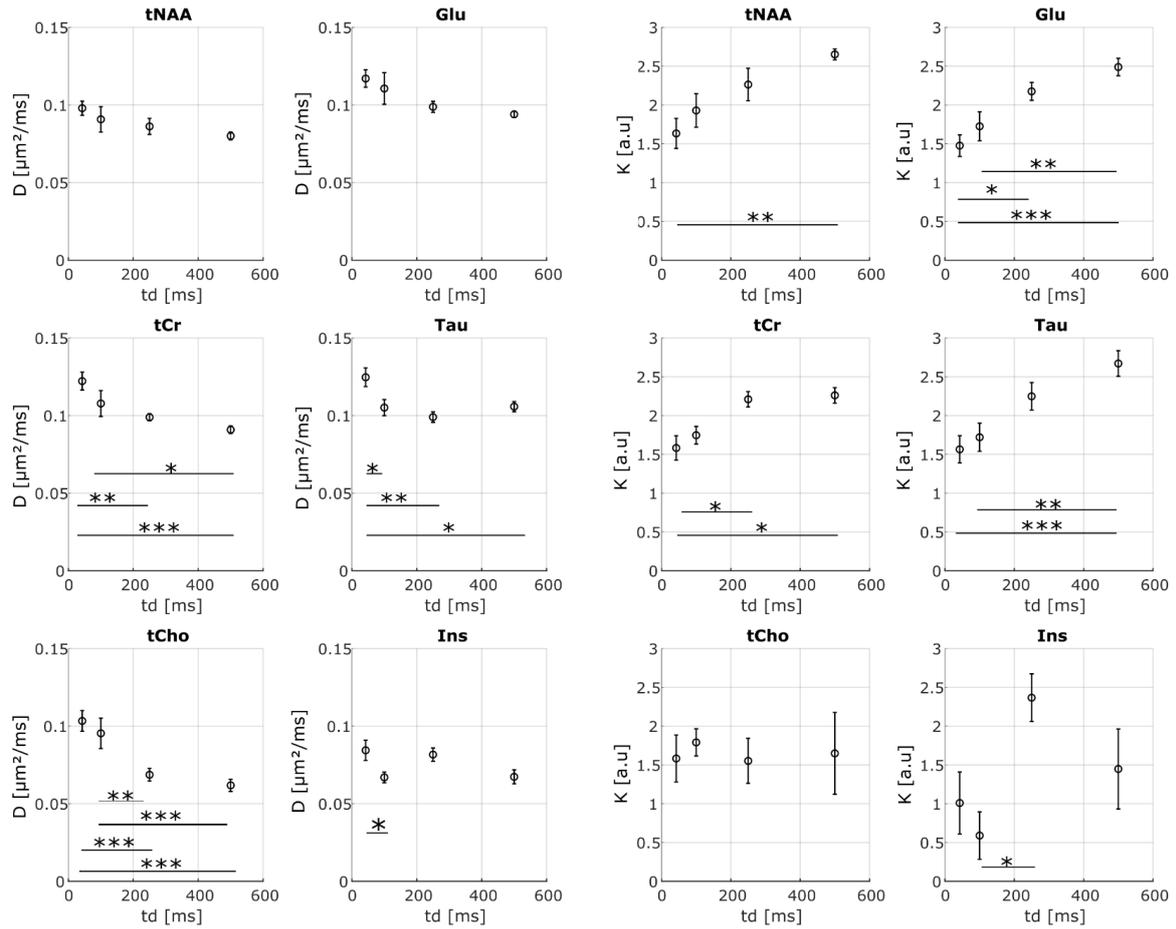

Fig.5: $D_M$ (left panel) and $K_M$ (right panel) averaged over seven mice as a function of diffusion time. For each metabolite, $D_M$ decreases and $K_M$ increases.

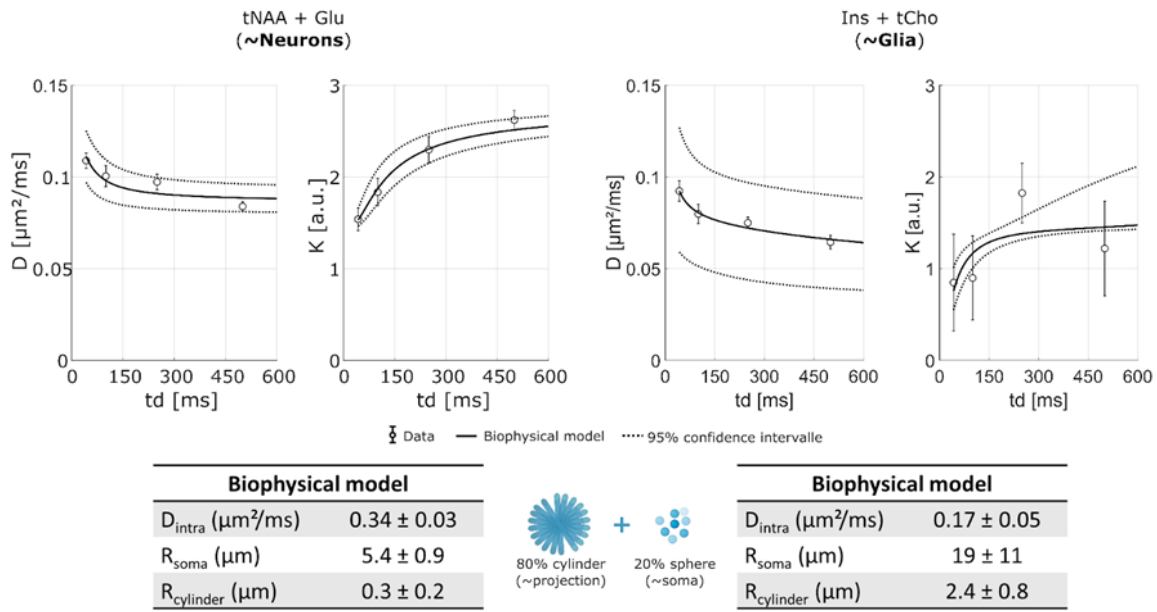

Fig.6: Fitting of a biophysical model to predominantly intraglial (upper panel) or preferentially intraneuronal (lower panel) metabolite data. The biophysical model is a very simple model composed of 80% isotropically distributed cylinders and 20% spheres, representing projections and soma, respectively.

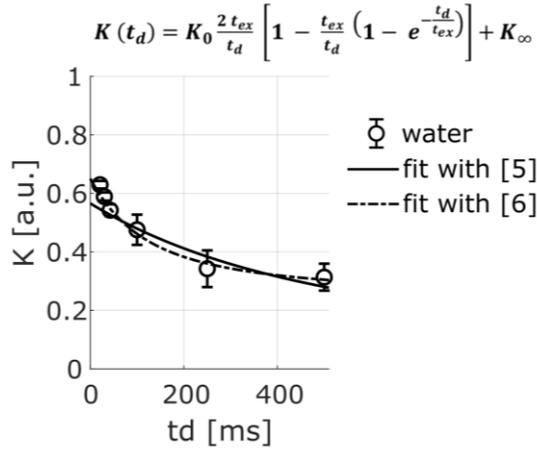

| $K(t_d)$ | [5] | [6] |
|---|---|---|
| | $K_0 = 0.57 \pm 0.18$ | $K_0 = 0.41 \pm 0.76$ |
| | $K_\infty = 0$ (fixed) | $K_\infty = 0.24 \pm 0.75$ |
| | $t_{ex} = 193 \pm 238$ ms | $t_{ex} = 46 \pm 412$ ms |
| BIC | -16 | -21 |
| cAIC | -15 | -19 |

Fig.7: Fit of the Kärger model (equation [5]) and the constant-modified Kärger model (equation [6]) to the $K_W$ data for $t_d \in [42.5; 500$ ms$]$, where $D_W(t_d)$ is nearly constant.

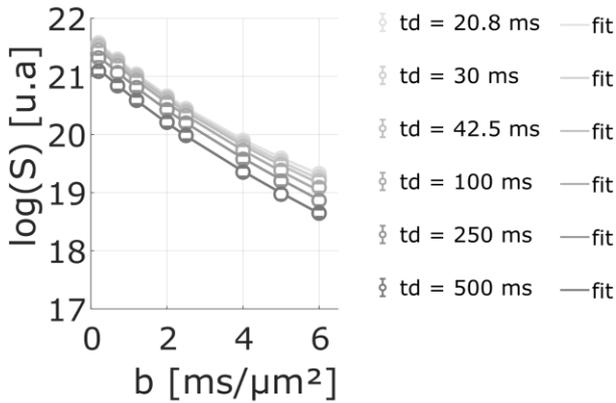

| | NEXI |
|---|---|
| | $f = 0.64 \pm 0.52$ |
| | $D_i = 1.56 \pm 1.28$ µm²/ms |
| | $D_e = 0.75 \pm 1.10$ µm²/ms |
| | $t_{ex} = 1.72 \pm 4.91$ ms |
| BIC | -283 |
| cAIC | -279 |

Fig.8: Fit of the water signal (log(S)) as a function of $b$ up to 6 ms/µm² with the NEXI/SMEX model.